\documentstyle[12pt,axodraw]{article}
\newcommand{\Fh}[2]{\,{}_#1F_#2}
\newcommand{\Fs}[3]{\!\!\left[\begin{array}{c}#1\,;\\#2\,;\end{array}#3\right]}

\newcommand{\Fup}[2]{\Fs{#1}{#2}{\frac{q^2}{m^2}}}

\newcommand{\Ffp}[2]{\Fs{#1}{#2}{\frac{q^2}{4m^2}}}
\newcommand{\Ffo}[2]{\Fs{#1}{#2}{\frac{4m^2}{4m^2-q^2}}}

\newcommand{\Li}[1]{\,{\rm Li}_{#1}}
\begin{document}
\thispagestyle{empty}
\onecolumn
\date{}
\vspace{-1.4cm}
\begin{flushleft}
{DESY 96-068 \\}
{JINR E2-96-62 \\}
{hep-th/9606018\\}
June,~1996

\end{flushleft}
\vspace{1.5cm}

\begin{center}

{\LARGE {\bf
   Connection between Feynman integrals
     having different values of the space-time
     dimension}
}

\vfill
{\large
 O.V.Tarasov \footnote{E-mail:
 $tarasov@ifh.de$ \\
 {~On leave of absence from JINR, 141980 Dubna (Moscow Region),
 Russian Federation.}}}

\vspace{2cm}
Deutsches Electronen-Synchrotron DESY \\
Institut f\"ur Hochenergiephysik IfH, Zeuthen\\
Platanenallee 6, D--15738 Zeuthen, Germany
\vspace{1cm}
\end{center}

\vfill

\begin{abstract}
A systematic algorithm for obtaining recurrence relations for
dimensionally regularized  Feynman integrals w.r.t.  the space-time
dimension $d$ is proposed. The relation between $d$ and $d-2$
dimensional integrals is given in terms of a differential operator
for which an explicit formula can be obtained for each Feynman
diagram. We show how the method works for one-,  two- and three-loop
integrals. The new  recurrence relations w.r.t. $d$ are complementary
to the recurrence  relations which derive from the method of
integration by parts. We find that the problem of the
irreducible numerators in  Feynman integrals can be naturally solved
in the framework  of the proposed generalized recurrence relations.
\end{abstract}

\vfill
\newpage

\setcounter{footnote}{0}
\section{Introduction}
Many  phenomena in high energy, solid state  and statistical physics
can be described by quantum field theoretical models considered
not only in four dimensional space-time but also in two-, three- and
other  space time dimensions \cite{Books}. In many cases perturbation
theory is the basic tool for calculating different physical
quantities. The same Feynman diagrams occur in different models with
different space-time dimension $d$. This parameter may be considered
as a regularization parameter \cite{HV} different from the value of
the physical space-time dimension. Usually one needs to set up a
Laurent expansion of the diagram w.r.t.  $\varepsilon=(2l-d)/2$,
with $2l$ being the dimension of the physical space -time in  the
problem under consideration. The coefficients of the expansion
are different for different $l$. The most advanced methods
were developed for the evaluation of $d=4-2\varepsilon$ dimensional
integrals. For example, the standard package MINCER \cite{LTV}
for calculating massless three-loop diagrams is now widely available
and has been used in many multiloop calculations.

Another reason to consider $d$ different from its physical value
was given in \cite{Bender}. There, a new approach for obtaining
nonperturbative information from a quantum field theory by expanding
Green's functions as a series in powers of $d$ was proposed.
In light of this investigation one can try to formulate
approximate methods for calculating {\it individual} Feynman
diagrams at some asymptotic values of $d$, for example as
$|d| \rightarrow \infty$ or as $|d| \rightarrow 0$.

One of the most powerful methods for evaluating Feynman diagrams is
the method of integration by parts \cite{CT}. In this approach
dimensionally regularized Feynman integrals are  considered
as functions of the exponents of the scalar propagators.
Integration by parts gives relations connecting  integrals
with some exponents changed by $\pm1$, very similar to
the relations for the contiguous hypergeometric functions.
This is not a big surprise since it has been  known for some time
\cite{Regge} that Feynman amplitudes belong to the
class of hypergeometric functions. In many cases, integrals
are just proportional to the well-studied hypergeometric
functions \cite{BoDa},\cite{BFT},\cite{Buza}.

Using the recurrence relations, a variety of Feynman integrals
can be reduced to the restricted set of  so called ``master
integrals''. The proof of completeness for the set of recurrence
relations obtained by  integration by parts and the problem of a
systematic algorithm how to use them for an arbitrary diagram
are open questions. Any additional information in this
respect may be useful for solving both problems.

 In the present paper we propose a systematic formulation of the
recurrence relations w.r.t. $d$. We shall show that these relations
cannot be obtained by the method of integration by parts  and
therefore should be considered as an important
addition to the Chetyrkin-Tkachov (CT) recurrence relations.
As a concrete example, we demonstrate that CT recurrence
relations with the recurrence relation w.r.t. $d$ compose all
possible recurrence relations for the considered integral.
 New recurrence relations are not of purely academic interest.
We demonstrate how one can calculate Feynman integrals using
these relations. We also rederive some useful relations for one-loop
diagrams with arbitrary number of external legs. The relations
connecting integrals with different $d$ may be also useful
for calculating integrals with  $d \neq 4$.

We show how tensor integrals can be represented in terms of
integrals with the changed $d$.
This representation allows us to write Feynman integrals with
irreducible numerators as a combination of scalar integrals
having different values of $d$. Thus, the solution of the generalized
system of recurrence relations automatically leads to the
solution of the problem of irreducible numerators.

It turns out that the new recurrence relations are especially
useful  in the new method for the  momentum expansion of the scalar
Feynman integrals proposed in \cite{sme}.

The paper is organized as follows. In Sect.2, we present the main
ideas of our method. First, using the parametric representation we
derive relation for arbitrary scalar integrals with different $d$.
Then we show how tensor integrals can be expressed in terms of
combinations of scalar integrals having different values of the
space-time dimension. In the framework of this
approach the solution to the problem of irreducible
numerators in Feynman integrals is proposed. An algorithm for
obtaining new generalized recurrence relations including integrals
with different $d$ as well as  different exponents of the propagators
is developed.

In Sect.3,  explicit relations connecting integrals with shifted $d$
for some one-, two- and three-loop integrals  are derived.

 In Sect.4, analogous relations  for the one-loop integrals
with an arbitrary number of external legs and arbitrary
powers of  propagators are given. We reproduced already
known results for these integrals  and obtained
more general new ones.

In Sect.5, we demonstrate how one can calculate integrals  explicitly
by solving recurrence relations w.r.t. the dimension of space-time.

 In Sect.6, the new recurrence relations are used to calculate the
$\varepsilon=(6-d)/2$ expansion for the two-loop self-energy diagrams
from the expansion developed at $\varepsilon=(4-d)/2$.


\section{Relations for integrals with different $d$}


The subject of our consideration will be dimensionally regularized
scalar Feynman integrals. An arbitrary scalar $L$ loop integral can
be written as
\begin{equation}
G^{(d)}(\{s_i\},\{m_s^2\})=\prod_{i=1}^{L}\int d^dk_i \prod^{N}_{j=1}
P^{\nu_j}_{\overline{k}_j,m_j},
\label{arbdia}
\end{equation}
where
\begin{equation}
P^{\nu}_{k,m}=\frac{1}{(k^2-m^2+i \epsilon)^{\nu}},~~~~~~
\overline{k}_j^{\mu}=~
 \sum^L_{n=1} \omega_{jn}k^{\mu}_n+ \sum _{m=1}^E
 \eta_{jm}q_m^{\mu},
\end{equation}
$q_m$ are external momenta,  $\{s_i\}$ is a set of scalar invariants
formed from $q_m$, $N$ is the number of lines,
$E$ is the number of external legs, $\omega$ and $\eta$ are matrices
of incidences of the diagram with the matrix elements being $\pm1$ or
$0$ (see, for example, Ref. \cite{IZ}).

To find the desired relation we shall use the parametric
representation of the integral, which can be found  in the literature
$\cite{IZ}$, $\cite{BM}$. For an arbitrary scalar Feynman integral in
$d$ dimensional space-time we have:
\begin{equation}
G^{(d)}(\{s_i\},\!\{m_s^2\})=i^L \left ( \frac{\pi}{i} \right)^
{\frac{dL}{2}}
\! \prod^{N}_{j=1} \frac{i^{-\nu_j}}{\Gamma(\nu_j)}
\! \int_0^{\infty} \!\!\! \ldots \!\! \int_0^{\infty}\frac{d \alpha_j
 \alpha^{\nu_j-1}_j}{ [ D(\alpha) ]^{\frac{d}{2}}}
 e^{i[\frac{Q(\{s_i\},\alpha)}{D(\alpha)}
 -\sum_{l=1}^{N}\alpha_l(m_l^2-i\epsilon)]},
\label{DQform}
\end{equation}
where   $D(\alpha)$
and $Q(\{s_i\},\alpha)$ are homogeneous polynomials in $\alpha$ of
degree $L$ and $L+1$, respectively. They can be represented
 as sums over trees and two-trees of the graph
(see, for example, \cite{BS}):
\begin{eqnarray}
D(\alpha)&=&\sum_{
\begin{array}{l}
over\\trees
\end{array}} ( \prod_{\begin{array}{l}over\\chords
\end{array}} ...\alpha_j...), \nonumber \\
&&\nonumber \\
Q(\{s_i\},\alpha)&=&\sum_{\begin{array}{l}
 over \\2-trees
 \end{array}} ~~ \prod_{\begin{array}{l}
 over\\ chords
 \end{array}}
\ldots \alpha_j \ldots ~~ (\sum_{
\begin{array}{l}{ over~comp.}\\ { of~2- tree}
\end{array}} q~)^2.
\label{trees}
\end{eqnarray}
These polynomials are characteristic functions of the topology
of the diagram and of its subgraphs.
Since $D$ and $Q$ will play an important role in the rest of the
present paper we remind the reader of the  definitions of the trees
and two-trees for   connected diagrams.
Any connected subdiagram of the diagram $G$ containing all the
vertices of $G$ but is free of cycles (loops) is called a tree of $G$.
Similarly, a two-tree is defined as any subdiagram of $G$
containing all the vertices of the original diagram, but is free of
cycles, and consisting of exactly two connected components. Finally,
a chord of a tree (two-tree) is defined as any line not belonging to
this tree (two-tree). Examples of $D$ and $Q$ for some
particular diagrams will be given in the next sections.

In the case when the $\nu_j$'s  do not depend on $d$, one can see
from ($\ref{DQform}$), that $d$ enters the integrand
in a rather simple way. Only the exponent of $D(\alpha)$
linearly depends on the dimension of space-time.

 In order to find the relation between integrals in different
dimensions $d$ we have to assume at first, that all scalar
propagators in ($\ref{arbdia}$) have different masses. Next, we
introduce the polynomial differential operator
\begin{equation}
D\left( \frac{\partial}{\partial m_j^2} \right),
\end{equation}
which is obtained from $D(\alpha)$ by substituting
$\alpha_j \rightarrow \partial_j \equiv \partial / \partial m_j^2$.
The  application of the operator $D(\partial)$
to the integral ($\ref{DQform}$)
gives $D(\alpha)$ in the numerator of the integrand:
\begin{equation}
 D (\partial )
e^{-i \sum \alpha_lm^2_l} \rightarrow D(\alpha) (-i)^L
e^{-i \sum \alpha_lm^2_l}.
\end{equation}
The resulting integral is proportional
to the same integral with $d$ changed to  $d-2$:
\begin{equation}
G^{(d-2)}(\{s_j\},\{m_s^2\})=\left(-~\frac{1}{\pi}\right)^L
D(\partial)
G^{(d)}(\{s_j\},\{m_s^2\}).
\label{connection}
\end{equation}
After having performed the differentiation we identify masses
with the one's of  the original integral.

We may include  tensor integrals into  our consideration using standard
methods \cite{IZ}, \cite{BS}. To each line one introduces an
auxiliary vector $a_j$ and then differentiates w.r.t. these vectors.
The parametric representation for a tensor integral with
products of $n_j$ vectors corresponding to the $j$-th line reads:
\begin{eqnarray}
&&
\prod_{i=1}^{L} \int d^dk_i
\prod_{j=1}^{N}  P^{\nu_j}_{\overline{k}_j,m_j}
\prod_{l=1}^{n_1} \overline{k}_{1\mu_l}  \ldots
\prod_{s=1}^{n_N} \overline{k}_{N\lambda_s}=
i^{L} \left( \frac{\pi}{i} \right)^{\frac{dL}{2}}
 \prod^{N}_{j=1}
 \frac{i^{-\nu_j-n_j}}{\Gamma(\nu_j)} \nonumber \\
&&~~\times
       \prod_{r=1}^{n_1} \frac{\partial}{\partial a_{1\mu_r}    }
\ldots \prod_{s=1}^{n_N} \frac{\partial}{\partial a_{N\lambda_s}}
\int_0^{\infty} \!\!\! \ldots \!\! \int_0^{\infty} \frac{  d \alpha_j
 \alpha^{\nu_j-1}_j}{ [ D(\alpha) ]^{\frac{d}{2}}}
 e^{i[\frac{Q(\{ \overline{s}_i \},\alpha)}
 {D(\alpha)}-\sum_{l=1}^{N}\alpha_l(\overline{m}_l^2-i\epsilon)]}
 \left|_{a_j=0} \right.,
\label{tensor}
\end{eqnarray}
where
\begin{equation}
\overline{m}_l^2=m_l^2+\frac{a_l^2}{4\alpha_l^2},
\end{equation}
$\overline{s}_i$ are scalar invariants formed from vectors
$\overline{q}_i$:
\begin{equation}
\overline{q}_i=q_{i}+ \sum_{j} \epsilon_{ij}
a_{j} \frac{1}{2 \alpha_j},
\label{auxvect}
\end{equation}
the $q_i$'s are the external momenta incoming
at a vertex $i$
 and $\epsilon$ is the incidence
 matrix defined as:
\[
\epsilon_{ij}= \left\{ \begin{array}{r@{\quad: \quad }l}
 +1& {\rm if~the ~oriented ~line} ~j~ {\rm points~ away~ from~
 the~ vertex}~ i\\
 -1& {\rm if~ the~ oriented~ line~} j {\rm~ points ~toward~
 the~ vertex}~ i\\
  0& {\rm if~ the ~ line~} j {\rm~ does~
  not~ contain~ the~ vertex~} i.
\end{array}  \right.
\]
Differentiation w.r.t. $a_i$ will produce external momenta and metric
tensors $g_{\mu \nu}$ times some polynomials $R_s(\alpha)$ divided by
$D(\alpha)$ to some power in the integrand. The polynomials
$R_s(\alpha)$ have to be converted into  operators $R_s(\partial)$
and the powers of $D(\alpha)$  are absorbed into the redefinition of
$d$. In this way any tensor integral  will be  expressed as a sum
over a set of tensors formed from external vectors and metric tensors
 multiplied by a combination of scalar integrals with the shifted
 value of $d$.
At the one-loop level such a representation was already
proposed in \cite{Andrey}.

Tensor integrals in momentum space can be written in terms of scalar
ones without direct appeal to the parametric representation
($\ref{tensor}$). The procedure described above may be derived from
the following formula:
\begin{equation}
\prod_{i=1}^{L} \int d^dk_i
\prod_{j=1}^{N}  P^{\nu_j}_{\overline{k}_j,m_j}
\prod_{r=1}^{n_1} \overline{k}_{1\mu_r}  \ldots
\prod_{s=1}^{n_N} \overline{k}_{N\lambda_s}=
T(q,\partial, {\bf d^+}) \prod_{i=1}^{L} \int d^dk_i
\prod_{j=1}^{N}  P^{\nu_j}_{\overline{k}_j,m_j},
\label{equation12}
\end{equation}
where the tensor operator $T$ is:
\begin{eqnarray}
&&
T(q,\partial,{\bf d^+})=\frac{e^{-iQ(\{ \overline{s}_i \},\alpha) \rho}
 }{i^{n_1+ \ldots +n_N}}
\nonumber \\
&&~~~~~~~~~~
\times
\!\prod_{r=1}^{n_1} \frac{\partial}{ \partial a_{1\mu_r}} \!\!\
\ldots \prod_{s=1}^{n_N} \left. \frac{\partial}{ \partial a_{N\lambda_s}}
e^{i[ Q(\{ \overline{s}_i \},\alpha)-\sum_{l=1}^N \frac{a_l^2}
{4\alpha_l} D(\alpha)] \rho }
\right|_{\begin{array}{l} a_j=0\\ \alpha_j=i \partial_j\\
\rho=(-\frac{1}{\pi})^L{\bf d^+}
\end{array}}.
\label{Ttensor}
\end{eqnarray}
Here ${\bf d^+}$ is the operator shifting the value of the space-time
dimension of the integral by two units: ${\bf d^+}G^{(d)}=G^{(d+2)}$.
As before we have to assume that at the beginning  all propagators
have different masses and after applying the operator $T$ to the
integral one should set masses equal to the required ones.

The representation ($\ref{equation12}$) for tensor integrals
may give a  solution of the problem of irreducible numerators,
i.e. the appearance of scalar invariants in the numerator
which are absent in the scalar propagators.
This kind of integrals can be expressed in terms of scalar
integrals without numerators but with the changed space-time
dimension. Therefore the combined set of recurrence relations,
i.e. relations with the changed $d$ and those obtained from
integration by parts,  should be used to reduce all integrals
to the set of scalar master integrals. In \cite{sme} we presented a
method for the  small momentum expansion of multiloop scalar
integrals based on $d$-recurrences.
This method from the very beginning does not produce
integrals with irreducible numerators although those appear
in traditional methods for the small momentum expansion.
In principle irreducible numerators can be regarded
as propagators raised to negative powers and the corresponding
integrals can be considered as object in a more complicated
class of integrals with additional new  denominators.
In our approach one remains in the same class of functions,
satisfying generalized recurrence relations i.e. recurrence relations
derived from the method of integration by parts
plus relations connecting integrals with different $d$.

Having representation ($\ref{equation12}$) at hand, we can now state
the procedure for obtaining new, generalized recurrence relations.
The  starting identity:
\begin{equation}
\prod_{i=1}^{L} \int d^dk_i
\frac{\partial }{\partial k_{r\mu}} \left\{
\left( \sum_l x_l \overline{k}_{l\mu} \right)
\prod^{N}_{j=1}
P^{\nu_j}_{\overline{k}_j,m_j} \right\} \equiv 0,
\label{ibpm}
\end{equation}
where $x_l$ are arbitrary constants, written in this form
turns out to be  rather convenient for the derivation.
After performing the differentiation one would usually express scalar
products with integration
momenta in terms of invariants that occur in the denominator of the
propagators. At this point we propose to change the derivation and
write all or only some of integrals containing scalar products with
loop momenta  as a combination of integrals with changed $d$.
In this way we  produce many
relations including integrals with changed exponents of scalar
propagators  and changed values of the  space-time dimension.
Combining the different relations (or choosing in a proper way $x_l$
in ($\ref{ibpm}$) ) one can try to find the most optimal set of
relations for the reduction of the concrete class of integrals to
the set of basic "master" integrals.
In fact, the method of integration by parts corresponds to some
specific representation of scalar products in ($\ref{ibpm}$).
Our derivation is more general and it includes integration
by parts method \cite{CT} as a particular case.

We expect that solutions of the generalized recurrence relations and
those obtained by the method of integration by  parts will be of the
same complexity. The fact that in case we know the explicit
result for an integral in terms of hypergeometric functions,
$d/2$ appears in the same manner as any exponent of the propagators,
can be considered as a confirmation of this statement.
However further investigation of this problem is needed.

Several remarks are in order. All relations which connect  integrals
with shifted $d$ are valid for  arbitrary
momenta and masses and also for their  real and imaginary parts.
 We want to stress here that
in general scalar products in ($\ref{connection}$)
must be considered as  independent variables. One  cannot use
any restrictions valid for some specific values of $d$.

All relations obtained with the help of the parametric representation
may  be profitably used in the frame of momentum as well as
configuration space.

The minimal change of the space-time parameter is two.
From concrete examples \cite{BoDa}-\cite{Buza}, when the result
is known in terms of hypergeometric functions,
one can observe that changing $d$ by $\pm 2$ we obtain
contiguous functions and therefore may hope to find
relations between integrals. The change of  $d$ by $\pm 1$
shifts the parameters of the hypergeometric functions
by $\pm 1/2$ producing functions which does not belong
to the class of contiguous  functions. In general there are
no  relations between those  functions.

Application of the differential operators to the integral will
increase the powers of the denominators. To simplify  the r.h.s. of
($\ref{connection}$) the CT recurrence relations  can be used.
In general every integral in $d-2$ dimensions
can be reduced to combinations of a rational function of scalar
products of external momenta, masses and $d$ and rational multiples
of master integrals $I_j^{(d)}(\{s_i\},\{m_s^2 \})$:
\begin{equation}
G^{(d-2)}(\{s_i\},\{m_s^2 \})=\sum_{j} C_j(\{s_i\},\{m_s^2 \},d)
I_j^{(d)}(\{s_i\},\{m_s^2 \}).
\label{equation15}
\end{equation}
The set of basic integrals in $d-2$ dimensions can be expressed in
terms of the same integrals in $d$ dimensions:
\begin{equation}
I_k^{(d-2)}(\{s_i\},\{m_s^2 \})=\sum_{j} B_{kj}
(\{s_i\},\{m_s^2 \},d)I_j^{(d)}(\{s_i\},\{m_s^2 \}).
\label{equation16}
\end{equation}
This relation can be inverted and therefore we will
get a representation of an arbitrary  $d$ dimensional
basic scalar integrals in terms of $d-2$ dimensional ones.
Such a representation may be useful for some practical calculations.
The most evident practical application of the relation connecting
integrals with different $d$: having the $\varepsilon$ expansion
for some particular $d_0$ we can find similar expansion in
$d_0 \pm 2l$ ($l$ integer) dimensions.

The proposed relations can be used, for example, in the evaluation of
massless propagator type $2-2\varepsilon$ dimensional integrals
using the package MINCER  \cite{LTV} written for $4-2\varepsilon$
dimensional integrals.  The polynomials $Q(\{s_i\},\alpha),D(\alpha)$
can be easily constructed by means of a computer program for
any particular integral \cite{OVT}.

\section{Examples}

In this section  several illustrative
examples of the new recurrence relations will be presented.
We start with
the one-loop propagator type diagram with massive particles:
\begin{equation}
I^{(d)}_{\nu_1 \nu_2}(q^2,m_1^2,m_2^2)=\int  \frac{d^d k_1}
{[i\pi^{d/2}]} P_{k_1,m_1}^{\nu_1} P_{k_1-q,m_2}^{\nu_2}.
\label{i1ab}
\end{equation}
In this case $D(\alpha)=\alpha_1+\alpha_2$ and therefore
\begin{eqnarray}
I^{(d-2)}_{\nu_1 \nu_2}(q^2,m_1^2,m_2^2)
 \!&=&\!-( \partial_1 +\partial_2)
  I^{(d)}_{\nu_1 \nu_2}(q^2,m_1^2,m_2^2) \nonumber \\
\!&=&\!-\nu_1 I^{(d)}_{\nu_1+1~ \nu_2}(q^2,m_1^2,m_2^2)
  - \nu_2 I^{(d)}_{\nu_1~ \nu_2+1}(q^2,m_1^2,m_2^2).
\label{equation18}
\end{eqnarray}
Here and in the following $\alpha_i$ is assigned  to the line
(or propagator)  with mass $m_i$.
We can get another  recurrence relation connecting integrals with
different $d$ following the prescription given in Sect.2. From the
identity:
\begin{equation}
\int d^d k_1 \frac{\partial}{\partial k_{1\mu}} \left[
(k_1+q)_{\mu} P^{\nu_1}_{k_1,m_1}P^{\nu_2}_{k_1-q,m_2}
\right] \equiv 0,
\end{equation}
we obtain (omitting for the moment arguments of
$I^{(d)}_{\nu_1 \nu_2}$):
\begin{eqnarray}
&&\nu_1\!\int\!\frac{ d^d k_1}{[i\pi^{d/2}]}
(qk_1) P_{k_1,m_1}^{\nu_1+1} P_{k_1-q,m_2}^{\nu_2}
\nonumber \\
&&~~~~~
\!=\left(\frac{d}{2}-\nu_1\right)I_{\nu_1\nu_2
}^{(d)}-\nu_2I_{\nu_1-1~ \nu_2+1}^{(d)}-\nu_1 m_1^2
I_{\nu_1+1~\nu_2}
 ^{(d)}+\nu_2(q^2-m_1^2)I_{\nu_1 ~\nu_2+1}^{(d)}.
\label{rec1}
\end{eqnarray}
The integral with the scalar product $(qk_1)$ can be written as a
scalar integral with shifted $d$. According to ($\ref{trees}$)
and ($\ref{auxvect}$)
the function $Q$  with auxiliary vectors $a_1$, $a_2$, which are
related to the lines with masses $m_1$, $m_2$, respectively, is:
\begin{equation}
Q(\{ \overline{s} \},\alpha)=\left( q+\frac{a_1}{2\alpha_1}-
\frac{a_2}{2\alpha_2}  \right)^2 \alpha_1 \alpha_2.
\end{equation}
Formulae  ($\ref{equation12}$) and ($\ref{Ttensor}$) for this
case yield the relation:
\begin{equation}
\int \frac{d^d k_1}{[i\pi^{d/2}]}~(qk_1)~ P_{k_1,m_1}^{\nu_1+1}
P_{k_1-q,m_2}^{\nu_2}= \nu_2 q^2  I^{(d+2)}_{\nu_1+1~ \nu_2+1}.
\label{vect1}
\end{equation}
Inserting ($\ref{vect1}$) into ($\ref{rec1}$)
we obtain the desired identity:
\begin{equation}
\nu_1 \nu_2 q^2 I^{(d+2)}_{\nu_1+1~\nu_2+1}
-\!\left(\frac{d}{2}-\!\nu_1 \right)\!I_{\nu_1 \nu_2}^{(d)}\!
+\nu_2 I_{\nu_1-1 ~\nu_2+1}^{(d)}+\nu_1 m_1^2I_{\nu_1+1~\nu_2}^{(d)}
\!-\nu_2 (q^2\!-m_1^2)I_{\nu_1~\nu_2+1}^{(d)}\!\equiv 0.
\label{obl1}
\end{equation}
In addition to the above relations two more relations can be obtained
from the traditional method of integration by parts:
\begin{eqnarray}
&&2\nu_2 m_2^2 I^{(d)}_{\nu_1~\nu_2+1}
+\nu_1 I^{(d)}_{\nu_1-1~ \nu_2+1}
+\nu_1 (m_1^2+m_2^2-q^2) I^{(d)}_{\nu_1+1~ \nu_2}
-(d-2\nu_2-\nu_1) I^{(d)}_{\nu_1 \nu_2}=0,\nonumber  \\
&& \nonumber \\
&&\nu_1 I^{(d)}_{\nu_1+1~ \nu_2-1}\!-\!\nu_2  I^{(d)}_{\nu_1
-1~\nu_2+1}\!-\!\nu_1(m_1^2\!-m_2^2\!+q^2) I^{(d)}_{\nu_1+1~\nu_2}
\!-\!\nu_2
(m_1^2\!-\!m_2^2\!-\!q^2) I^{(d)}_{\nu_1~ \nu_2+1}
\nonumber \\
&& \nonumber \\
&&~~~~~~~~~~~~~~~~~~~~~~~~~~~~~~~~~~~~~~~~~~~~~~~~~~~~~~~~~~
+(\nu_2\!-\!\nu_1) I^{(d)}_{\nu_1 \nu_2}=0.
\label{CTferm}
\end{eqnarray}
For the simplest case, at $m_1=m_2=0$ and $\nu_1=\nu_2=1$ from
($\ref{equation18}$), ($\ref{CTferm}$) we readily get:
\begin{equation}
I^{(d-2)}_{11}(q^2,0,0)=-2I^{(d)}_{21}(q^2,0,0)=\frac{2(d-3)}{q^2}
I^{(d)}_{1 1}(q^2,0,0).
\end{equation}
This formula can be easily verified from the explicit result for
$I^{(d)}_{\nu_1 \nu_2}(q^2,0,0)$.

At $m_1=0,~m_2=m$, the relations connecting integrals
$I^{(d)}_{\nu_1 \nu_2}(q^2,0,m^2)$ with different $d$ are:
\begin{eqnarray}
&&\nu_1 \nu_2 q^2I^{(d+2)}_{\nu_1+1~ \nu_2+1}(q^2,0,m^2)
-\left(\frac{d}{2}-\nu_1 \right)I_{\nu_1 \nu_2}^{(d)}(q^2,0,m^2)
+\nu_2 I_{\nu_1-1~ \nu_2+1}^{(d)}(q^2,0,m^2) \nonumber \\
&&~~~~~~~~~~~~~~~~~~~~~~~~~~~~~~~~~~~~~~~~~~~~~~~~~~~~~~~~~~~~~~~~
-\nu_2 q^2 I_{\nu_1~\nu_2+1}^{(d)}(q^2,0,m^2) \equiv 0, \nonumber \\
&&I^{(d-2)}_{\nu_1 \nu_2}(q^2,0,m^2)+
\nu_1 I^{(d)}_{\nu_1+1~\nu_2}(q^2,0,m^2)+
\nu_2 I^{(d)}_{\nu_1~ \nu_2+1}(q^2,0,m^2) \equiv 0.
\label{myferm}
\end{eqnarray}
The integral $I^{(d)}_{\nu_1 \nu_2}(q^2,0,m^2)$
is proportional to the
Gauss hypergeometric  function \cite{BoDa}:
%
%
\begin{equation}
I^{(d)}_{\nu_1 \nu_2}(q^2,0,m^2)=(-1)^{\nu_1+\nu_2}
 \frac{\Gamma(\nu_1+\nu_2-\frac{d}{2}) \Gamma(\frac{d}{2}-\nu_1) }
 {(m^2)^{\nu_1+\nu_2-\frac{d}{2}} \Gamma(\frac{d}{2}) \Gamma(\nu_2)}
\Fh21\Fup{\nu_1,\nu_1+\nu_2-\frac{d}{2}}{ \frac{d}{2} }.
\label{BoDa}
\end{equation}
As is well known there are fifteen relations of Gauss between
contiguous functions $_2F_1$. Substituting ($\ref{BoDa}$)
into  ($\ref{CTferm}$) one can find
correspondence between the CT
recurrence relations  and only six  relations of  Gauss.
The reason is obvious - in the CT relations the third parameter
of  $_2F_1$ in ($\ref{BoDa}$) does not change and therefore all
corresponding relations for contiguous functions cannot
be reproduced. If we include into consideration also identities
($\ref{myferm}$) we cover all fifteen relations,
though in principle, to reduce the integrals
$I^{(d)}_{\nu_1 \nu_2}(q^2,0,m^2)$, with integer  $\nu_1$ and $\nu_2$
to two boundary integrals, it is enough to know only the CT relations.

  Now we consider two-loop bubble integrals with three different
masses:
\begin{equation}
J^{(d)}_{\nu_1 \nu_2 \nu_3 }
=\frac{1}{\pi^d} \int \!\! \int {d^d k_1~d^d k_2}~P_{k_1,m_1}^{\nu_1}
P_{k_1-k_2,m_2}^{\nu_2}P_{k_2,m_3}^{\nu_3}.
\label{i2abg}
\end{equation}
The function $D(\alpha)$ for this integral is:
\begin{equation}
D(\alpha)=\alpha_1 \alpha_2 +\alpha_1 \alpha_3 +\alpha_2 \alpha_3,
\end{equation}
and hence,
\begin{equation}
J^{(d-2)}_{n l r}=
(\partial^2_{12} +\partial^2_{13} +\partial^2_{23})
J^{(d)}_{\nu_1 \nu_2 \nu_3}.
\label{dtodm2}
\end{equation}
In the above formula and in what follows we use the shorthand
notation:
$$
\partial^N_{i_1 \ldots i_N}=\prod_{j=1}^N \frac{\partial}
{\partial m^2_{i_j}}.
$$
Let's take  for simplicity $\nu_1=\nu_2=\nu_3=1$.
In this particular case the relation connecting $J^{(d)}_{111}$
with different $d$ first was found  in \cite{magic}. Exploiting
the relation obtained by the method of integration by parts:
\begin{equation}
\Delta~ \partial_1 J^{(d)}_{1 1 1}
=\frac12(d-3)(\partial_1 \Delta) J^{(d)}_{1 1 1}
+(d-2)J^{(d)}_{0 1 1}+\frac{(d-2)}{4m_1^2}\left[(\partial_3 \Delta)
 J^{(d)}_{1 0 1}
 +(\partial_2 \Delta)  J^{(d)}_{1 1 0} \right],
\end{equation}
where
\begin{equation}
\Delta
=m_1^4+m_2^4+m_3^4-2m_1^2m_2^2-2m_1^2m_3^2-2m_2^2m_3^2,
\end{equation}
and similar relations for $\partial_2 J^{(d)}_{111}$,
$\partial_3 J^{(d)}_{111}$, from ($\ref{dtodm2}$)
we reproduce the relationship   between  integrals with
different $d$ given in \cite{magic}:
\begin{equation}
\Delta J^{(d-2)}_{111}=-(d-2)(d-3)J^{(d)}_{111}
+\frac{1}{2} \Gamma^2 \left( 2-\frac{d}{2}
\right)
[s_1s_2 ~\partial_3 \Delta
 +s_1s_3~ \partial_2 \Delta
 +s_2s_3~ \partial_1 \Delta],
\end{equation}
where $s_i=m_i^{d-4}$.

One can easily obtain another  recurrence relation for
the integral ($\ref{i2abg}$) with different $d$. From the identity
\begin{equation}
\int \!\! \int d^dk_1~d^dk_2~\left( \frac{\partial}{\partial k_{1\mu}}
+\frac{\partial}{\partial k_{2\mu}} \right)
~\left[ k_{1\mu}~
P_{k_1,m_1}^{\nu_1}
P_{k_1-k_2,m_2}^{\nu_2}P_{k_2,m_3}^{\nu_3} \right] \equiv 0,
\end{equation}
keeping the scalar product $(k_1k_2)$ untouched, we get:
\begin{equation}
\frac{\nu_3}{\pi^d} \!\!\int \!\! \int\!\!d^d k_1~d^dk_2~
(k_{1}k_2)~
P_{k_1,m_1}^{\nu_1}
P_{k_1-k_2,m_2}^{\nu_2}P_{k_2,m_3}^{\nu_3+1}
+\!\nu_1 m_1^2~J^{(d)}_{\nu_1+1~ \nu_2 \nu_3}\!-\left(
\!\frac{d}{2}-\!\nu_1\!\right)\!J^{(d)}_{\nu_1 \nu_2 \nu_3}
\equiv 0.
\label{oblique2}
\end{equation}
 The first integral in  ($\ref{oblique2}$) can be expressed in terms
of integrals with another value of the space-time
dimension $d$ by using formula ($\ref{Ttensor}$) with:
\begin{equation}
Q(\{\overline{s} \}, \alpha)=\frac14\left(\frac{a_1}{\alpha_1}-
\frac{a_2}{\alpha_2}-\frac{a_3}{\alpha_3} \right)^2
\alpha_1\alpha_2\alpha_3.
\end{equation}
Here again vectors  $a_i$ correspond to  lines with mass $m_i$. In
order to obtain the integral with the scalar product $(k_1k_2)$ in
the integrand one has to differentiate w.r.t. $a_1$ and $a_3$ which
leads to the relation:
\begin{equation}
\frac{1}{\pi^d}~\int d^dk_1~d^dk_2
(k_{1}k_2)~
P_{k_1,m_1}^{\nu_1}
P_{k_1-k_2,m_2}^{\nu_2}P_{k_2,m_3}^{\nu_3+1}
=\frac12 \nu_2 d J^{(d+2)}_{\nu_1~ \nu_2+1~ \nu_3+1}.
\label{k1k2}
\end{equation}
Substituting ($\ref{k1k2}$) into ($\ref{oblique2}$)
we obtain:
\begin{equation}
\nu_2\nu_3 d J^{(d+2)}_{\nu_1~ \nu_2+1~ \nu_3+1}+2\nu_1 m_1^2~J^{(d)}
_{\nu_1+1~ \nu_2 \nu_3}-(d-2\nu_1)J^{(d)}_{\nu_1 \nu_2 \nu_3}
\equiv 0.
\end{equation}
This identity was used to evaluate  the
coefficients in the small momentum expansion of the two-loop
master diagram with all masses equal by the method proposed
in \cite{sme}.

At the three-loop level, bubble diagrams in general have the
topology shown in Fig.1,  where  each line
corresponds to a scalar propagator with an arbitrary exponent.
\vspace{0.2cm}
\begin{center}\begin{picture}(160,70)(0,0)
\SetWidth{0.75}
\SetScale{0.6}
\CArc(80,50)(50,0,180)
\CArc(80,50)(50,180,360)
\Line(80,50)(45,85)
\Line(80,50)(115,85)
\Line(80,50)(80,0)
\Text(40,-18)[]{${\rm Fig.1. ~~ Three-loop ~bubble~ diagram}$}
\Text(47,68)[]{$1$}
\Text(83,28)[]{$2$}
\Text(12,28)[]{$3$}
\Text(35,35)[]{$4$}
\Text(63,35)[]{$5$}
\Text(53,17)[]{$6$}
\end{picture}\end{center}
\vspace{0.8cm}
The relation between $d$ and $d-2$ dimensional three-loop
vacuum integrals with all masses arbitrary is:
\begin{eqnarray}
&&G^{(d-2)}(\{m^2_s\})
  =-\frac{1}{\pi^3} \left( \partial^3_{123}
 +\partial^3_{124}+\partial^3_{126}+\partial^3_{135}
 +\partial^3_{136}+\partial^3_{145}+\partial^3_{146}
 +\partial^3_{156}
 \right. \nonumber \\
&&~~~~~~~~~ \left.
 +\partial^3_{234}+\partial^3_{235}
 +\partial^3_{245}+\partial^3_{246}+\partial^3_{256}
 +\partial^3_{345}+\partial^3_{346}+\partial^3_{356}
 \right )G^{(d)}(\{m^2_s\}).
\label{D3loop}
\end{eqnarray}
The numbering  of the masses corresponds to the numbering of the
lines  in Fig.1.

Three-loop master bubble integrals encountered in the small momentum
expansion of the QED photon propagator   were
studied  in \cite{David54}. Every  integral   in this case
can be expressed in terms of three basic structures.
The only nontrivial combination of two integrals taken as one of such
structure was:
\begin{eqnarray}
&&m^{3d-12}B_4^{(d)}=-\frac{1}{4}(d-2)(d-3)
 \int\!\! \int\!\! \int \!\! \frac{d^dk_1~d^dk_2~d^dk_3}{[i\pi^{d/2}
 \Gamma(3-\frac{d}{2})]^3}
~P_{k_2,m}P_{k_3,m}P_{k_2-k_3,0}  \nonumber\\
&&  \nonumber\\
&&~~~~~~~\times
\left[ P_{k_1,m}~P_{k_1-k_3,0}~P_{k_1-k_2,0}
 -P_{k_1,0}~P_{k_1-k_3,m}~P_{k_1-k_2,m} \right].
\end{eqnarray}
By using ($\ref{D3loop}$) and the method of integration by parts
we find the following relation for $B_4^{(d)}$:
\begin{eqnarray}
&&B_4^{(d-2)}=\frac{3(d-4)^2(d-5)(3d-14)(3d-16)}{16(d-6)^5}B_4^{(d)}
+\frac{33d^3-367d^2+1170d-864}
{8(d-4)^2(d-6)^5}  \nonumber \\
&&~~~~- \frac{7(8d^2-80d+195)}
{(d-4)^2(d-6)^5} ~
\frac{\Gamma(\frac{d}{2}-1)\Gamma^2(6-d)\Gamma(9-\frac{3d}{2})
}{\Gamma^2(3-\frac{d}{2})\Gamma(12-2d)}.
\label{B4}
\end{eqnarray}
We observe that the relation includes $B_4^{(d)}$ itself and two
additional terms   with a trivial $d$ dependence. The results for
other diagrams of the three-loop photon propagator look similar, i.e.
they exhibit  three terms of a structure like in ($\ref{B4}$).

Other useful  three-loop vacuum integrals were introduced in
\cite{AFMT}:
\begin{eqnarray}
&&\!\!\!\!\!\!
m^{3d-12}~D_3=\int \!\!\! \int \!\!\! \int \frac{d^d k_1 d^d k_2
d^d k_3}{[i \pi^{d/2} \Gamma(3-d/2)]^3}
P_{k_1,0}P_{k_2,0}P_{k_3,m}P_{k_1-k_3,0}
P_{k_1-k_2,m}P_{k_2-k_3,m},\\
&&\!\!\!\!\!\!m^{3d-12}~B_5=
\int \!\!\! \int  \!\!\! \int \frac{d^d k_1 d^d k_2 d^d k_3}{
[i \pi^{d/2} \Gamma(3-d/2)]^3}
P_{k_1,m}P_{k_3,m}P_{k_1-k_2,m}P_{k_2-k_3,0}.
\end{eqnarray}
These integrals occurred in the evaluation of the three-loop
QCD correction to the electroweak $\rho$ parameter.
The space-time recurrence relation for $D_3^{(d)}$  looks more
complicated than that for $B_4^{(d)}$ and it includes also
$B_5^{(d)}$:
\begin{eqnarray}
&&D_3^{(d-2)}=
-\frac{16(63d^3-832d^2+3622d-5176)}
{(d-4)^4(d-5)^2(d-6)^3}
-\frac{64\Gamma(\frac{d}{2}-1)
\Gamma(5-d)}{(d-4)^3(d-5)(d-6)^3~\Gamma(3-\frac{d}{2})} \nonumber \\
&& \nonumber \\
&&~~\times \left[1+\frac{(7d-32) ~\Gamma(\frac{d}{2}-1)
 \Gamma(7-\frac{3d}{2})} {3 ~\Gamma^2(3-\frac{d}{2})}
+\frac{(37d^2-350d+828)~\Gamma(5-d)\Gamma(7-\frac{3d}{2})}
{24(2d-9) ~\Gamma(3-\frac{d}{2}) \Gamma(9-2d)}\right]
\nonumber \\
&& \nonumber \\
&&~~~-~\frac{4(d-2)(d-3)(3d-8)(3d-10)}{9(d-4)(d-6)^3}B_5^{(d)}
-\frac{4(d-2)(d-3)(d-4)}{(d-6)^3}D_3^{(d)}.
\label{D3}
\end{eqnarray}
The corresponding relation for $B_5^{(d)}$ is  simpler  and reads:
\begin{equation}
B_5^{(d-2)}=-\frac{4(d-2)(d-3)(3d-8)(3d-10)}{9(d-4)(d-6)^3}
B_5^{(d)}-\frac{64(15d-52)}{9(d-4)^4(d-6)^3}.
\label{b5}
\end{equation}
It can be obtained from ($\ref{D3loop}$) by observing  that
$\partial_2 G^{(d)}=\partial_4 G^{(d)}=0$; which holds because
$B_5^{(d)}$ corresponds  to the diagram  Fig.1 with shrinked second
and fourth lines. For $B_5^{(d)}$ we found an explicit expression
in terms of hypergeometric functions $_3F_2$ and $_2F_1$ with the
argument $1/4$, satisfying relation ($\ref{b5}$).

Relations ($\ref{B4}$),($\ref{D3}$) and ($\ref{b5}$) can be used
for the computation of the  coefficients in the small
momentum expansion of Feynman diagrams
by the method proposed in \cite{sme}.

A more detailed  analysis of the generalized recurrence
relations for two- and three- loop diagrams will be given in
a future publication. Several examples of these recurrence
relations one can find also in \cite{sme}.
\section{Relations for n-point one-loop integrals}

In this section we consider  scalar one-loop integrals  depending
on $n-1$  external momenta:
\begin{equation}
I_n^{(d)}=
\int \frac{d^d q}{\pi^{{d}/{2}}} \prod_{j=1}^{n}
\frac{1}{[j]^{\nu_j}},
\end{equation}
where
\begin{equation}
[j]=(q-p_j)^2-m_j^2+i\epsilon ~~~{\rm for}~~{j<n}~~~{\rm and}
~~~[n]=q^2-m^2_n+i\epsilon.
\end{equation}
The corresponding diagram and the convention for
the momenta are given in Fig.2.
\begin{center}
\begin{picture}(300,100)(0,0)
\SetWidth{0.75}
\SetScale{0.7}
\ArrowLine(115,100)(185,100)
\ArrowLine(90,50)(115,100)
\ArrowLine(115,0)(90,50)
\ArrowLine(165,0)(115,0)
\ArrowLine(185,100)(205,57)
\DashCArc(155,50)(50,-90,5){3}
\ArrowLine(90,125)(115,100)
\ArrowLine(210,125)(185,100)
\ArrowLine(65,50)(90,50)
\ArrowLine(90,-25)(115,0)
\Text(110,80)[]{$q-p_2$}
\Text(162,60)[]{$q-p_{3}$}
\Text(55,61)[]{$q-p_1$}
\Text(58,20)[]{$q$}
\Text(103,10)[]{$q-p_{n-1}$}
\Text(100,-30)[]{Fig.2.~~One-loop diagram with $n$-legs}
\end{picture}
\end{center}
\vspace{1.5cm}
For the diagram  Fig.2
the  function $D(\alpha)$ is a polynomial linear in $\alpha_j$:
\begin{equation}
D(\alpha)=\sum_{j=1}^{n} \alpha_j,
\end{equation}
and hence
\begin{equation}
I_n^{(d-2)}=-\sum_{j=1}^{n} \partial_j I_n^{(d)}.
\label{muleg}
\end{equation}
To get rid of the derivatives in this relation we use the method of
integration by parts. Let's consider the identity:
\begin{equation}
\int d^dq \frac{\partial}{\partial q_{\mu}}
 \left [ \frac{x_nq_{\mu}+\sum_{i=1}^{n-1} x_ip_{i \mu}}
 {\prod_{j=1}^{n} [j]^{\nu_j} }\right] \equiv 0,
\end{equation}
that is valid for arbitrary $x_i$. Upon differentiation and
expressing scalar products in the numerator
in terms of  invariants $[j]$ in the denominator,  we get:
\begin{eqnarray}
&&\int d^dq \frac{1}{\prod_{i=1}^{n}[i]^{\nu_i}} \left \{
 x_n \left( d-\nu_n- \sum_{j=1}^{n} \nu_j \right) +
 \sum_{j=1}^{n-1}x_j \left(\nu_j-\nu_n+  [j] \sum_{k=1,k\neq j}^{n-1}
 \frac{\nu_k}{[k]} \right)
 \right. \nonumber \\
&&\nonumber \\
&&~~~~~~~~~~~~~~~~~~~~~\left. -[n]\sum_{j=1}^{n-1}\frac{\nu_j}{[j]}
 \sum_{i=1}^{n}x_i+\sum_{j=1}^{n}\nu_j
 \frac{R^{(n)}_j(\{x_i\})}{[j]} \right\}\equiv 0,
\label{mulegeq}
\end{eqnarray}
where $R^{(n)}_{j}(\{x_i\})
=\sum_{i=1}^{n}R^{(n)}_{ij}x_i~~$ with
\begin{eqnarray}
&& R^{(n)}_{ni}=-m_i^2-m_n^2+p_i^2, ~~~~~~~~~~~
 R^{(n)}_{in}=m_i^2-m^2_n-p_i^2 ~~~~{\rm for~ } i < n,   \nonumber \\
&&R^{(n)}_{ij}=m_i^2-m^2_n-p_i^2+2p_ip_j,~~
~~~~~~~~~~~~~~~~~~~~~~~~~~~~~~~~~{\rm for~ } i,j < n
\nonumber \\
&&R^{(n)}_{nn}=-2m^2_n.
\end{eqnarray}
The  integral  related to the last sum in ($\ref{mulegeq}$) can be
made proportional to the
r.h.s. of ($\ref{muleg}$). This  requires to find  $x_i$ such that
\begin{equation}
R^{(n)}_j(\{x_i\})=1,~~~~~j=1,\ldots,n.
\end{equation}
Imposing these conditions we obtain a system of $n$ equations
for $x_i$. The solution of the system reads
\begin{equation}
x_i=\frac12~ \partial_i \ln  \Delta_n  ~~~{\rm for }~~i< n,~~~~~~~
x_n=\frac{G_{n-1}}{2 \Delta_n},
\end{equation}
where $G_n$ is the Gram determinant

\[
G_n=\left| \begin{array}{llcl}
p_1p_1 &p_1p_2 &\cdots &p_1p_n \\
p_2p_1 &p_2p_2 &\cdots &p_2p_n \\
 \multicolumn{4}{c}\dotfill    \\
p_np_1 & p_np_2 &\cdots &p_np_n
\end{array}
 \right| , \]
and $\Delta_n$ is proportional to the  determinant
\begin{equation}
\Delta_n=-\frac{1}{2^n} \det( C ),
\end{equation}
of the $n \times n$ matrix $C$:
\begin{equation}
C_{ij}=m_{i-1}^2+m_{j-1}^2-(p_{i-1}-p_{j-1})^2,
\end{equation}
where it is assumed that $m_0=m_n$ and $p_0=0$. Substituting $x_i$
into ($\ref{mulegeq}$), we obtain the following relation
between  $d$ dimensional and $(d-2)$ dimensional integrals:
\begin{equation}
\left[ \sum_{j=1}^{n-1}(\nu_j-\nu_n) \left(\partial_j  \Delta_n
 \right)
 +(d-\nu_n-\sum_{j=1}^{n} \nu_j)G_{n-1} \right]
I_n^{(d)}=
\sum_{j=1}^{n} \left(\partial_j \Delta_n\right)
I_{n,j}^{(d-2)}+2 \Delta_n I_{n}^{(d-2)}.
\label{new1loop}
\end{equation}
The index $j$ of $I^{(d-2)}_{n,j}$ means that the factor
$1/[j]^{\nu_j}$ in the integrand must be changed into
$[j]/[j]^{\nu_j}$. For $\nu_j=1$ we obtain the simpler relation:
\begin{equation}
(d-n-1)G_{n-1}I_{n}^{(d)}=
\sum_{j=1}^{n} \left(\partial_j \Delta_n \right)
I_{n,j}^{(d-2)}+2\Delta_n I_{n}^{(d-2)}.
\label{bdk}
\end{equation}
For $d=6$, $n \geq 6$ and assuming that external momenta are four
dimensional the Gram determinant $G_{n-1}$ vanishes and hence
the term with $I_{n}^{(6)}$ drops out and we get
a relationship between $4$-dimensional integrals with  $n$ legs
and integrals with
$n-1$ legs.
Such a relation  was first  obtained in \cite{brown}
(see also \cite{melrose}).
When $d=6$ and $n=5$ the l.h.s. of ($\ref{bdk}$) is zero and
we arrive at a  formula for reducing $4$ - dimensional
pentagon integrals to  box integrals. This relation  was first
derived  in \cite{Halpern}.
For arbitrary $d$ and $n$  the equation ($\ref{bdk}$)
 was obtained by a different method in \cite{BDK}.
Eq. ($\ref{new1loop}$) as far as we know is new.

It is evident that by using ($\ref{connection}$) and integration
by parts one can also derive  relations similar to ($\ref{bdk}$)
for multiloop integrals.

Equation ($\ref{new1loop}$) can be used in the  reduction of
the one-loop
tensor integrals to scalar ones. In \cite{Andrey} an explicit
general formula for one-loop  tensor integrals was derived:
%
\begin{eqnarray}
\int { d^d q}
\frac{q_{\mu_1} \ldots q_{\mu_M}}{ \prod_{j=1}^{n} [j]^{\nu_j} }=
\sum_{
\begin{array}{l}
\lambda, k_1, \ldots ,k_{n-1} \\2\lambda+\sum {k_i}=M
\end{array}}
\frac{(-1)^{M-\lambda}}{2^{\lambda}}
\{ [g]^{\lambda}
[p_1]^{k_1} \ldots [p_{n-1}]^{k_{n-1}} \}_{\mu_1 \ldots \mu_M}
&&\nonumber \\
~~~~~~~~~~~~~~~~~~~~~~~~~~~~~~~~~~~~~~~\times
\int\frac{ d^{d+2M-2\lambda} q }
{ \pi^{M-\lambda}[n]^{\nu_n} }\prod_{j=1}^{n-1}
\frac{(\nu_j)_{k_j}}{[j]^{\nu_j+k_j}},
\label{andre}
\end{eqnarray}
where $(\nu)_k\equiv \Gamma(\nu+k)/\Gamma(\nu)$ is the Pochhammer
symbol. The shorthand notation
$\{ [g]^{\lambda}[p_1]^{k_1} \ldots [p_{n-1}]^{k_{n-1}} \}_{\mu_1
\ldots \mu_M}$
corresponds to the symmetrical (w.r.t. ${\mu_1 \ldots \mu_M}$)
tensor combination, each term of which is constructed from $\lambda$
metric tensors $g$, $k_1$ momenta $p_1$, $\ldots$, $k_{n-1}$ momenta
$p_{n-1}$. For example,
$$
\{gp_1\}_{\mu_1 \mu_1 \mu_3}=g_{\mu_1\mu_2}p_{1\mu_3}+
g_{\mu_1\mu_3}p_{1\mu_2}+g_{\mu_2\mu_3}p_{1\mu_1}.
$$
In ($\ref{andre}$) $\lambda, k_i \geq 0$, max $k_i=M$, max
$\lambda =[M/2]$ (integer part of $M/2$). For more details see
\cite{Andrey}. Thus, the procedure of calculating  one-loop diagrams
will be as follows: one should use first ($\ref{andre}$),
perform contractions of indices and then using ($\ref{new1loop}$)
to reduce all scalar integrals with the changed dimensions to
the $d=4-2\varepsilon$ dimensional set of   integrals.
After that scalar integrals should be reduced to the
set of master integrals by making use of recurrence relations
obtained by the method of integration by parts.

\section{Evaluation of one-loop integrals using
space-time recurrence relations}
With the help of equations ($\ref{equation15}$) and
($\ref{equation16}$) one can obtain  relations that include  only one
particular integral with different shifts in  $d$:
\begin{equation}
\sum_{k} B_{k} (\{s_i\},\{m_s^2 \},d)I^{(d-k)}(\{s_i\},\{m_s^2 \})=
I_0(\{s_i\},\{m_s^2 \},d),
\label{equation59}
\end{equation}
where $I_0(\{s_i\},\{m_s^2 \},d)$ is some explicitly known
expression. It can be eliminated from the equation giving rise to
higher order recurrence relations. For an explicit solution
equation ($\ref{equation59}$) looks  simpler
than the ones obtained by the method of integration by parts.
In this section we will demonstrate how the space-time recurrence
relations proposed in the present paper can be used for evaluating
integrals.

We consider the following one-loop propagator type integral:
\begin{equation}
I^{(d)}=m^{4-d} \int \frac{d^dk_1}{[i\pi^{d/2}]}~ P_{k_1,m}
P_{k_1-q,m}.
\end{equation}
By using ($\ref{connection}$) and the first relation
of ($\ref{CTferm}$) we obtain:
\begin{equation}
I^{(d-2)}=\frac{2(d-3) m^2}{(q^2-4m^2)}I^{(d)}
 -\frac{2 \Gamma(2-\frac{d}{2})  m^2 }{(q^2-4m^2)}.
\label{jrec}
\end{equation}
Introducing
\begin{equation}
\bar{I}^{(d)}=i^{-d} \left( \frac{m^2}{4m^2-q^2}\right)^{\frac{d}{2}}
\frac{\Gamma(d-2)}{\Gamma(\frac{d}{2}-1)} I^{d},
\end{equation}
we get the simpler equation:
\begin{equation}
\bar{I}^{(d-2)}=\bar{I}^{(d)}-\left(\frac{m^2}{4m^2-q^2}\right)
^{\frac{d}{2}}
 \frac{ \Gamma(d-3) \Gamma(2-\frac{d}{2})}
 {i^d \Gamma(\frac{d}{2}-1)}.
\label{jbar}
\end{equation}
Without loss of generality, one can  parametrise $d$  as
$ d=2l-2\varepsilon,$
where $l$ is an integer number and $\varepsilon$ is an arbitrary
small number. The inhomogeneous term in ($\ref{jbar}$) can be
absorbed by a  redefinition of $\bar{I}^{(d)}$:
\begin{equation}
\bar{I}^{(2l-2\varepsilon)}=\sum_{j=0}^{l}
\left(\frac{m^2}{4m^2-q^2}\right)^{j-\varepsilon}
\frac{\Gamma(2j-2\varepsilon-3)
\Gamma(2-j+\varepsilon)}{i^{2j-2\varepsilon} \Gamma(j-1-\varepsilon)}
+C^{l}_{\varepsilon},
\label{ceps}
\end{equation}
yielding the following very simple equation for $C^{l}_{\varepsilon}$:
\begin{equation}
C^{l}_{\varepsilon}=C^{l-1}_{\varepsilon}.
\end{equation}
Since $l$ in our case is an arbitrary integer we can conclude that
 $C^{l}_{\varepsilon}$ does not depend on $l$ at all. It can be
found, for example, by taking the limit $l\rightarrow \infty$, or
$|d| \rightarrow \infty$. Taking the limit $|d| \rightarrow \infty$
is quite a delicate matter. The larger $|d|$, the more divergent
becomes the integral.  In order to keep the regularization of the
integral, we have to  consider $\varepsilon$ as  complex  and $l$
large. The asymptotic value of the integral as $ l \rightarrow \infty$
can be found by the method of steepest descent.
It depends on the sign of $q^2$.
  From the parametric representation of the
integral $I^{(d)}$   one gets  for $q^2<0$  and ${ l \to\infty}$ :
\begin{equation}
I^{(d)}
\rightarrow  \Gamma\left( 2-l+\varepsilon \right)
\left(\frac{-\pi m^2}{q^2 l}\right)^{\frac12} \left(1-\frac{q^2}{4m^2}
\right)^{l-\varepsilon-\frac{3}{2}}(1+O(1/l)).
\end{equation}
At large $l$ and $q^2<0$, the sum in ($\ref{ceps}$) is convergent
which allows us to find $C^{l}_{\varepsilon}$:
\begin{equation}
C^{l}_{\varepsilon}=\frac{i^{2\varepsilon} \pi}{\sin \pi \varepsilon }
\frac{m^4}{\sqrt{- q^2}(4m^2-q^2)^{3/2}}-
\sum_{j=0}^{\infty}
\left(\frac{m^2}{4m^2-q^2}\right)^{j-\varepsilon}
\frac{\Gamma(2j-2\varepsilon-3)
\Gamma(2-j+\varepsilon)}{i^{2j-2\varepsilon} \Gamma(j-\varepsilon-1)}.
\end{equation}

Substitution of  $C^{l}_{\varepsilon}$ in the above formulae
yields:
\begin{eqnarray}
&&I^{(d)}=\frac{\sqrt{\pi} \Gamma \left(\frac{d}{2}-1 \right)
\Gamma \left(2-\frac{d}{2}\right)}{\Gamma \left(\frac{d-1}{2}\right)}
 ~\frac{m}{\sqrt{-q^2}} \left(1-\frac{q^2}{4m^2} \right)^{\frac{d-3}
 {2}}\nonumber \\
&&~~~~~~~~~~~~~~~~~~~~~~~~~~~~~~
+\Gamma \left(1-\frac{d}{2}\right) \frac{2m^2}{(4m^2-q^2)}
\Fh21\Ffo{1,\frac{d-1}{2}}{\frac{d}{2}}.
\end{eqnarray}
By using the formula for the analytic continuation of the
hypergeometric series we get the known result:
\begin{equation}
I^{(d)}=
\Gamma\left( 2-\frac{d}{2} \right) \Fh21\Ffp{1,2-\frac{d}{2}}
{\frac{3}{2}}.
\end{equation}

The example considered illustrates the  main ideas of how to use the
relation ($\ref{connection}$) for the evaluation of Feynman integrals.
The same method applies without modification
to more complicated cases. For example, to the  one-loop integral
($\ref{i1ab}$) with different masses at $\nu_1=\nu_2=1$ or to the
two-loop integral ($\ref{i2abg}$) with $nu_1=\nu_2=\nu_3=1$.
The recurrence relations in these cases are very similar to
($\ref{jrec}$).
We applied this technique also to some two- and three loop integrals.
The main difficulty encountered in these computations was to find
the asymptotic value for the integral for large $d$. An expansion at
large $d$ is frequently used in solid state physics and statistical
physics. Our experience shows that it can also be used for the
approximate evaluation of {\it individual} Feynman integrals.
Details of these calculations will be given in a future publication.


\section{The two-loop self-energy for the
$\phi^3$ model in $6-2\varepsilon$ dimensions}


The aim of this section is to show how $d$-recurrence
relations can be used to obtain the $\varepsilon$ expansion
of the nontrivial integrals in one dimension from
the known expansion in another dimension.

We consider two-loop self-energy diagrams which are encountered,
for example, in the  $\phi^3$ model with two massive and one massless
scalar fields \cite{fikub}. At the two loop level there are
two nontrivial diagrams, given in Fig.3, contributing to the
self-energy of the massless field.
\begin{center}
\begin{picture}(200,80)(0,-10)
\SetScale{1.5}
\CArc(30,25)(15,0,180)
\CArc(30,25)(15,180,0)
\DashLine(30,40)(30,10){2}
\DashLine(45,25)(55,25){2}
\DashLine(5,25)(15,25){2}
\Text(45,5)[]{$I_1^{(d)}$}
\Text(24,57)[]{$1$}
\Text(66,57)[]{$2$}
\Text(24,20)[]{$3$}
\Text(66,20)[]{$4$}
\Text(50,38)[]{$5$}
\DashLine(65,25)(75,25){2}
\CArc(90,25)(15,0,180)
\CArc(90,25)(15,180,0)
\DashLine(78,30)(102,30){2}
\DashLine(105,25)(115,25){2}
\Text(135,5)[]{$I_2^{(d)}$}
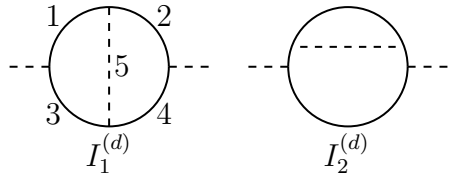
\Text(82,-16)[]{Fig.3. ~Two-loop scalar diagrams with massive loops}
\end{picture}
\end{center}
\vspace{0.5cm}
The solid lines in Fig.3 are related to massive particles
and the dashed lines to massless ones. The corresponding integrals
read:
\begin{eqnarray}
&&I_1^{(d)}=\int \!\! \int \frac{ d^d k_1 d^d k_2}{\left[i\pi^{
\frac{d}{2}} \Gamma(4-\frac{d}{2})\right]^2 }~
P_{k_1,m}P_{k_2,m}P_{k_1-q,m}P_{k_2-q,m}P_{k_1-k_2,0},
 \nonumber\\
&&I_2^{(d)}=\int \!\! \int \frac{d^d k_1 d^d k_2}{ \left[i
\pi^{\frac{d}{2}} \Gamma(4-\frac{d}{2})\right]^2 }
P_{k_1,m}^2P_{k_2,m}P_{k_2-q,m}P_{k_1-k_2,0}.
\end{eqnarray}

Using the  $\varepsilon$ expansion in $d=4-2\varepsilon $ dimensions
we will find the $\varepsilon$ expansion of these integrals up to
$O(\varepsilon)$ at $d=6-2\varepsilon$. This can be done in three
steps.

The first step consists of expressing the integrals $I_1^{(d)}$ and
$I_2^{(d)}$ in terms of  basic ones, chosen in \cite{BFT} to obtain
the $d \rightarrow 4$ limit for the two-loop photon propagator. In
fact two integrals from this basis are the integral $I_1^{(d)}$ itself
and its derivative w.r.t. the mass. The  remaining basic integrals
were the derivative w.r.t. the mass of the one-loop scalar integral
with two massive lines, its square and a one-loop vacuum integral
with a mass.

 Thus, at this stage we need to compute only the integral $I_2$.
 By using the  recurrence relations given in \cite{BFT} for integrals
 contributing to the two-loop photon propagator,  we get:
\begin{equation}
I_2^{(d)}=-\frac{(d-4)}{(d-3)}
I_1^{(d)}+\frac{m^2}{(d-3)}\left(1-\frac{q^2}{4m^2}\right)
\left(I_1^{(d)} \right)' +\frac{m^{2d-10}}{(d-3)(d-4)}H_1^{(d)},
\end{equation}
where the prime denotes differentiation w.r.t. $m^2$ and
$H_1^{(d)}$ is a basic  integral occurring in one-loop
calculations:
\begin{equation}
H_1^{(d)}=\Fh21\Ffp{1,3-\frac{d}{2}}{\frac32}=
\frac{1}{d-5}+\frac{(d-6)}{(d-5)}\left(1-\frac{q^2}{4m^2}\right)
H_1^{(d-2)}.
\end{equation}

The next step  is to find  relations between the basic
integrals $I_1^{(d)},~\left(I_1^{(d)} \right)'$
in $d$ and $d-2$ dimensions. Using Eq.($\ref{connection}$),
with:
\begin{equation}
D(\alpha)=\alpha_5(\alpha_1+\alpha_2+\alpha_3+\alpha_4)+
 (\alpha_1+\alpha_3)(\alpha_2+\alpha_4),
\end{equation}
($\alpha$'s are labeled as shown in Fig.3) we get a system of
equations in the form of ($\ref{equation16}$).
This system enables us to find the required relations:
\begin{eqnarray}
&& -3z(d-3)(d-4)^2 (3d-14)(3d-16)I_1^{(d)}\!=\frac{m^4}{4}(d-8)^2(d-6)
\!\! \left[-(d-6)(d-4)z^3 \right. \nonumber \\
&&~~\left. +12(d-4)(2d-11)z^2
+12(d-7)(3d-14)z-16(d-7)(2d-9)\right]I_1^{(d-2)}\nonumber \\
&&~~+\frac{m^6}{4}(d-8)^2(d-6)\left[2(d-4)z^3+2(3d-16)z^2-8(9d-41)z
\right.
\nonumber \\
&&~~\left.+32(2d-9)\right] \left(I_1^{(d-2)}\right)'
-\frac{2 m^{2d-10}}{(d-6)^2(4-z)} \left[(d-4)(23d^2-226d+552)z^3
\right. \nonumber \\
&&~~ -8(d-4)(19d^2-188d+462)z^2+16(22d^3-310d^2+1447d-2238)z
\nonumber \\
&&~~\left.+64(2d-9)(d-5)^2\right]\left(H_1^{(d)}\right)^2
+\frac{16 m^{2d-10}}{(d-6)^2(4-z)}
\left[(19d^2-182d+432)z^2 \right.\nonumber \\
&&~~\left. -(56d^2-552d+1344)z  +16(2d-9)(2d-11)\right] H_1^{(d)}
\nonumber \\
&&~~+\frac{16 m^{2d-10}}{(d-6)^2(d-4)(d-5)(4-z)}
     \left[(d-6)(d-4)^2z^2-2(d^3+3d^2-98d+288)z \right. \nonumber \\
&&~~~~~~~~~~~~~~~~~~~~~~~~~~~~~~~~~~~~~~~~~~~~~~~~~
\left. -8(d-5)(5d-28)(2d-9)\right],
\end{eqnarray}
\begin{eqnarray}
&&-3 z(d-3)(d-4)^2(3d-14)(3d-16)\left( I_1^{(d)}\right)'=
\frac{3m^2}{2}(d-4)(d-6)(d-8)^2 \nonumber \\
&&~~\times \left[(d-4)(d-5)z^2+(7d-34)(d-7)z-4(2d-9)(d-7)\right]
I_1^{(d-2)} \nonumber \\
&&~~+\frac{3m^4}{4}(d-4)(d-6)(d-8)^2
\left[(5d-24)z^2-4(7d-33)z+16(2d-9)\right]
\left( I_1^{(d-2)} \right)' \nonumber \\
&&~~+\frac{24(d-4)m^{2d-12}}{(4-z)(d-6)^2}\left[(d-4)(7d^2\!-\!70d
+174)z^2\! -(30d^3-424d^2\!+1986d-3084)z \right. \nonumber \\
&&~~\left. -8(2d-9)(d-5)^2\right]\left(H_1^{(d)}\right)^2
-\frac{96(d-4)m^{2d-12}}{(4-z)(d-6)^2}\left[(d-4)(d-6)z
\right.
\nonumber \\
&&~~
\left. -4(2d-9)(2d-11)\right] H_1^{(d)}
+\frac{48(7d-32)(d-6)z-192(2d-9)(5d-28)}{m^{12-2d}(4-z)(d-6)^2},
\end{eqnarray}
where $z=q^2/m^2$.

The final step is to perform the $\varepsilon$ expansion
up to a constant term.
It is remarkable that the most complicated integrals
$I_1^{(d-2)}$ and $\left(I_1^{(d-2)}\right)'$ at $ d=6-2\varepsilon$
give contribution starting from the first order
in $\varepsilon$  and therefore may be disregarded
in the considered approximation.
To calculate $I_1^{(d)}$ and $I_2^{(d)}$ up to a constant term
in $\varepsilon$ we need to expand $H_1^{(d)}$ to second order.
The required expansion  is:
\begin{eqnarray}
&&H_1^{(6-2\varepsilon)}=1+\varepsilon \left[2+\frac{1}{a} \ln(v)
\right] \nonumber \\
&&~~~~~+\varepsilon^2 \left[4+\frac{\pi^2}{6a}+\frac{1}{2a} \ln^2(v)
+\frac{2}{a}\ln(v)-\frac{1}{a}\ln^2\left(\frac{1+a}{2}\right)
-\frac{2}{a} \Li2 \left(\frac{1+a}{2}\right)\right],
\end{eqnarray}
where
\begin{equation}
a^2=\frac{q^2}{q^2-4m^2},~~~~~~~ v=\frac{1-a}{1+a},
{\rm~~~~~ and~~~~}\Li2(x)=-\int_0^x \frac{\ln(1-y)dy}{y},
\end{equation}
from which we obtain
\begin{eqnarray}
m^{4\varepsilon-2}I_1^{(6-2\varepsilon)}&=&
\frac{3-a^2}{6\varepsilon^2(1-a^2)}
 +\frac{31-7a^2}{12\varepsilon(1-a^2)}
+\frac{19}{24}-\frac{1}{12a^2}+\frac{47}{6(1-a^2)}
-\frac{2h(a)}{3a(1-a^2)} \nonumber \\
&~&-\left( \frac{5}{48}-\frac{1}{3a}-\frac{7}{24a^2}
+\frac{1}{48a^4}-\frac{1}{3(1-a)} \right) \ln^2 (v)
\nonumber \\
&& \nonumber \\
&~&
+\frac{1}{12}\left(\frac{43}{a}-\frac{1}{a^3}+
\frac{48a}{(1-a^2)} \right) \ln(v), \\
&& \nonumber \\
&& \nonumber \\
m^{4\varepsilon-2} I_2^{(6-2\varepsilon)}&=&
- ~\frac{a^2}{18\varepsilon^2(1-a^2)}-\frac{15+28a^2}
{108\varepsilon(1-a^2)}
 +\frac{389}{324}-\frac{1}{36a^2}-\frac{1207}{648(1-a^2)}
\nonumber \\
&& ~~\nonumber \\
&~&-\frac{1-3a^2}{18a(1-a^2)}
~h(a)
+\frac{1}{144} \left(10+\frac{4}{a}-\frac{1}{a^2}
 -\frac{1}{a^4}-\frac{8}{(1-a)} \right)
\ln^2(v) \nonumber \\
&& ~~\nonumber \\
&~&+ \left(
 \frac{5}{54a}-\frac{1}{36a^3}-\frac{43a}{54(1-a^2)}
 \right)\ln(v),
\end{eqnarray}
with
$$h(a)=
  \ln^2
 \left( \frac{1+a}{2}\right) -\frac{1}{\varepsilon} \ln(v)
-\zeta(2)
+2~ \Li2 \left( \frac{1+a}{2} \right).
$$

In the considered case it is  possible to obtain an explicit result
in terms of hypergeometric functions and hence   to perform the
$\varepsilon$ expansion using directly the exact formulae. However,
from our experience with the  $\varepsilon$ expansion at
$d=4-2\varepsilon$, we  may say that this is a quite formidable and
errorprone task. Thus, having done it once,  it is much simpler
to use the $d$- recurrence relations in order to get the result in the
other dimension. In cases when the analytic result is not known,
it might be useful to choose the most convenient $d$ from the point
of view of the $\varepsilon$  expansion and then to transform the
result to the required space-time dimension. Especially this procedure
maybe  helpful if  one encounters integrals
with infrared as well as ultraviolet divergences.

\section{Conclusions}

We proposed a new type of recurrence relations for  Feynman integrals.
 The addition  of the  space-time   dimension  to the set of
recurrence parameters for Feynman integrals is quite natural and it
 extends the set of recurrence relations obtained from the
 integration by parts method. We expect that within  the framework
 of the extended set of recurrence relations it will be possible to
 formulate rather effective algorithms for the computation of Feynman
 diagrams. The problem with irreducible numerators in Feynman
 integrals finds a natural interpretation in the frame of generalized
 recurrence relations. We demonstrated that a direct solution of
 the new recurrence relations
  is possible. Finding solutions of such relations would considerably
simplify if it would be possible to develop an  effective algorithm for
the asymptotic expansion  at $|d| \rightarrow  \infty$.
  At the same time the expansion at $|d| \rightarrow \infty$ could
be used as an approximation for the integral at finite $d$. This kind
of expansion is well known in solid state and statistical physics.
  Basically, in the present publication we formulated general ideas
  about new techniques which we intend to  apply later to the
  calculation of some particular classes of Feynman integrals.

\vspace{0.6cm}
{\Large {\bf Acknowledgments }}
\vspace{0.4cm}

I am grateful to  D.V. Shirkov and S.V. Mikhailov for their
proposal to perform the  evaluation of the integrals discussed
in section 6,  which became a starting point for the
present investigation. I am also grateful to J. Fleischer
and F. Jegerlehner for useful discussions and for carefully reading
the manuscript.

Financial supports from BMBF and from the Russian Basic Research
Foundation (grant N 93-02 14428) are gratefully acknowledged.
I want to thank also the Aspen Center for Physics (where part of this
work was done) for the warm hospitality  and the financial support.

\end{document}